\documentclass[a4paper]{jpconf}
\usepackage{graphicx}
\usepackage{amsmath}
\usepackage{subfigure}
\usepackage{dcolumn}
\usepackage{bm}
\usepackage{amssymb}
\usepackage{latexsym}
\usepackage{epsf}
\usepackage{graphics}
\usepackage{amssymb}
\usepackage{amsfonts}
\usepackage[mathscr]{eucal}
\usepackage{amsmath}
\usepackage{enumerate}
 \usepackage{hhline}
\usepackage{array}
\usepackage{tabularx}

\newcommand{\dd}{\mathrm{d}}

\newcommand{\uc}{\mathrm{c}}

\newcommand{\mean}[1]{\left\langle #1 \right\rangle}
\newcommand{\mpl}{m_{_\mathrm{Pl}}}

\newcommand{\ie}{{\it i.e.}~}

\newcommand{\etc}{{\it etc.}}

\newcommand{\bear}{\begin{array}}  \newcommand{\eear}{\end{array}}
\newcommand{\bef}{\begin{figure}}  \newcommand{\eef}{\end{figure}}
\newcommand{\bec}{\begin{center}}  \newcommand{\eec}{\end{center}}

\newcommand{\lkk}{\left[}  \newcommand{\rkk}{\right]}

\newcommand{\order}{{\cal O}}

\newcommand{\cl}{\mathrm{cl}}
\newcommand{\ini}{\mathrm{in}}

\newcommand{\df}{\delta\varphi}

\begin{document}
\title{Stochastic Inflation in Compact Extra Dimensions}

\author{Larissa Lorenz}

\address{Institute of Mathematics and Physics, Centre for
  Cosmology, Particle Physics and Phenomenology, Louvain
  University, 2 Chemin du Cyclotron, 1348 Louvain-la-Neuve, Belgium}

\ead{larissa.lorenz@uclouvain.be}

\begin{abstract}
While moving down the potential on its classical slow roll trajectory, the inflaton field is subject to quantum jumps, which take it up or down the potential at random. In ``stochastic inflation'', the impact of these quantum jumps is modeled by smoothing out the field over (at least) Hubble-patch sized domains and treating fluctuations on smaller scales as noise. The inflaton thus becomes a stochastic process whose values at a given time are calculated using its probability distribution. We generalize this approach for non-canonic kinetic terms of Dirac Born Infeld (DBI) type and investigate the resulting modifications of the field's trajectory. Since models of DBI inflation arise from string-inspired scenarios in which the scalar field has a geometric interpretation, we insist that field value restrictions imposed by the model's string origin must be respected at the quantum level.
\end{abstract}

\section{Introduction}
Recent advances in string theory have lead to the construction of explicit models of inflation in which the scalar field $\phi$ is, for example, associated with the position of a $(p+1)$-dimensional hypersurface (a so-called D$p$-brane) embedded into the ten-dimensional string background spacetime~\cite{Dvali:1998pa}. For $p=3$, the brane's world volume corresponds to our four-dimensional Universe, while the six extra dimensions are compactified on a suitable internal manifold~\cite{Kachru:2003sx}. It is hoped that string theory can bring us closer to establishing inflation's theoretical foundations, while the cosmological parameters derived from ``brane inflation'' may be the only potentially observable consequences of string theory for decades to come. Due to the D3's Dirac Born Infeld (DBI) kinetic term, these models provide a ``stringy'' realization of $k$-inflation~\cite{Garriga:1999vw}, giving rise to a rich and non-standard phenomenology.

While inflationary predictions for the properties of the cosmic microwave background evidently apply inside our observable patch of the Universe, the stochastic approach to inflation~\cite{Starobinsky:1994bd} seeks to understand the behavior of the quantum inflaton field on \emph{much} larger scales. In this work~\cite{Lorenz:2010vf}, we extend the stochastic treatment applied to standard inflation in Refs.~\cite{Martin:2005ir,Martin:2005hb} to DBI scenarios. There, the impact of the inflaton's stochastic character on scales beyond the Hubble patch was (in the Gaussian approximation) entirely expressed in terms of integrals of the Hubble parameter $H(\phi)$ and its derivatives. Here, we find that the influence of the modified kinetic term is readily absorbed into the same formalism if the so-called Lorentz factor $\gamma(\phi)$~\cite{Alishahiha:2004eh}, which measures the dynamics' deviation from the usual $\dot{\phi}^{2}/2$ (with $\gamma\approx1$ in the standard and $\gamma\rightarrow\infty$ in the ``ultrarelativistic'' case), is treated on an equal footing with $H(\phi)$.

\section{Stochastic inflation}
\subsection{The standard case}
The notion of the inflaton as a stochastic process $\varphi[\xi]$ 
arises as follows: the field 
 $\phi(\bm x,t)$ is coarse-grained on a given (large) scale $L$, meaning that a spatially smooth field $\varphi(t)$ is constructed from $\phi$'s long-wavelength Fourier modes only, while fluctuations $\delta\phi(\bm x,t)$ on smaller scales are interpreted as stochastic ``noise''. Explicitly, this decomposition reads
\begin{equation}
\phi(\bm x,t)=\varphi(t)+\int\frac{{\rm d}^3{\bm k}}{(2\pi)^{3/2}}\,
\Theta\left(k-L^{-1}\right)\,
\lkk a_{\bm k}\,\delta\phi_{\bm k}(t)\,\e^{-i\bm k\cdot \bm x}+
a_{\bm k}^\dag\,\delta\phi_{\bm k}^\ast(t)\,\e^{i\bm k\cdot \bm x} \rkk\,,
\label{eq:separation}
\end{equation}
where $L^{-1}=\varepsilon a H$ with $\varepsilon$ a small parameter. The smoothed $\varphi(t)$ then obeys a Langevin-type equation, which is  obtained from the slow-roll\footnote{These read $\dot{\varphi}\approx-V'/(3H)$ and $H^{2}\approx(\kappa/3)\,V$, where $\kappa\equiv8\pi/\mpl^{2}$.} Friedmann and Klein Gordon equations plus a noise term $\xi(t)$ due to the small-scale fluctuations (a prime denotes $\varphi$-derivatives),
\begin{equation}\label{eq:Langevin}
\dot{\varphi}=-\frac{2}{\kappa}\,H'
+\frac{H^{3/2}}{2\pi}\,\xi(t)\, ,\qquad\textnormal{where}\quad\mean{\xi(t)}=0\,,\quad \mean{\xi(t)\,\xi(t')}=\delta(t-t')\,.
\end{equation}
In Ref.~\cite{Martin:2005ir}, this equation was solved using a perturbative expansion in the noise, \ie by writing $\varphi(t)=\varphi_\cl(t)+\delta\varphi_{1}(t)+\delta\varphi_{2}(t)+\dots$, with $\delta\varphi_{1}(t)\propto\order{(\xi)}$, $\delta\varphi_{2}(t)\propto\order{\left(\xi^{2}\right)}$ \etc~By sorting the resulting terms on both sides of Eq.~(\ref{eq:Langevin}) by orders of $\xi$, evolution equations for $\delta\varphi_{1}$ and $\delta\varphi_{2}$ are identified, which can be solved by a time integration of $H$, its $\varphi$-derivatives and the noise. While the explicit time-dependence of $\xi(t)$ is unknown, its stochastic properties (\ref{eq:Langevin}) can be used to obtain the mean values $\mean{\varphi_{1}^{2}}$ and $\mean{\varphi_{2}}$ in terms of $\varphi$-space integrals of $H, H'$ \etc~only. These mean values enter into the probability density function (PDF) $P_\uc(\varphi)=\mean{\delta(\varphi-\varphi[\xi])}$, which describes the probability for the stochastic process $\varphi[\xi]$ to assume a given value $\varphi$ at time $t$ inside a single coarse-grained domain. In the Gaussian approximation, one has
\begin{equation}
\label{eq:Pc} 
P_{\rm c}(\varphi)=\frac{1}{\sqrt{2\pi\mean{\df_1^2}}}\,
\exp\left[-\frac{(\varphi-\varphi_\cl-\mean{\df_2})^{2}}
{2\mean{\df_1^2}}\right]\,,
\end{equation}
where $\varphi_\cl$ is the inflaton's value on the \emph{classical slow-roll} trajectory. Hence, since $\mean{\varphi_{1}^{2}}$ and $\mean{\varphi_{2}}$ are functions of $\varphi_\cl$ (and the initial field value $\varphi_\ini$ at the onset of inflation), one can use them to calculate the \emph{quantum-corrected} mean $\mean{\varphi}=\varphi_\cl+\mean{\delta\varphi_{2}}$ along with the standard deviation. 
Ref.~\cite{Martin:2005ir} carried out this calculation for several model classes, illustrating how quantum jumps significantly modify the classical trajectory in some cases.

Let us add two further remarks: first, the PDF of Eq.~(\ref{eq:Pc}) considers $\varphi[\xi]$ inside one $L^{3}$ volume. However, since $L\propto aH$ itself depends on the inflaton's classical value inside it, different domains should be ``weighted'' with respect to their size. This is achieved by calculating a volume-corrected distribution $P_{\mathrm{v}}(\varphi)$, which gives a $\mean{\varphi}_{\mathrm{v}}\neq\mean{\varphi}$~\cite{Martin:2005ir}. 
Second, the perturbative expansion evidently only holds if the noise-induced corrections become smaller and smaller at higher orders in $\xi$. In Ref.~\cite{Martin:2005hb}, this condition was quantified and used as an intrinsic criterion for the regime of validity of the above ansatz.

\subsection{The DBI case}
We showed~\cite{Lorenz:2010vf} that the results of Refs.~\cite{Martin:2005ir,Martin:2005hb} are straightforwardly generalized to the case of string-inspired DBI inflation, provided the coarse-graining scale in Eq.~(\ref{eq:separation}) is adapted and the DBI Langevin equation is used. It can be shown that the DBI analogue of Eq.~(\ref{eq:Langevin}) is obtained by replacing $H'\rightarrow H'/\gamma$ in the classical drift term\footnote{This can be shown using the ``DBI slow-roll'' equations $\dot{\varphi}\approx -V'/(3H\gamma)$ and $H^{2}\approx(\kappa/3)V$.}, where the Lorentz factor is given by $\gamma(\varphi)=\left[1+4H'^{2}/(\kappa^{2}\,T)\right]^{1/2}$. (For the DBI domain size, one has $L^{-1}_{_{\mathrm{DBI}}}=\varepsilon a\gamma H$.) Here, the so-called warping function $T(\varphi)$ --like the potential $V(\varphi)$-- is \emph{a priori} a free function but in a concrete model of brane inflation, both $V$ and $T$ are derived from the 10d string background~\cite{Dvali:1998pa,Kachru:2003sx}. Using the same perturbative expansion in $\xi$, one can again solve for $\mean{\delta\varphi_{1}^{2}}$ and $\mean{\delta\varphi_{2}}$ in terms of integrals along $\varphi$'s (DBI-)classical trajectory, which read 
\begin{eqnarray}
\mean{\delta\varphi_{1}^{2}}
&=&\frac{\kappa}{2}\left(\frac{H'}{2\pi\gamma}\right)^{2}
\int_{\varphi_{\mathrm{cl}}}^{\varphi_{\mathrm{in}}}\dd\psi\,
\left[\frac{H(\psi)\gamma(\psi)}{H'(\psi)}\right]^{3}\,,\label{eq:df1}\\ 
\mean{\delta\varphi_{2}}&=&\frac{\left(H'/\gamma\right)'}{2\left(H'/\gamma\right)}
\mean{\delta\varphi_{1}^{2}}+\frac{H'/\gamma}{4\pi\mpl^{2}}
\left[\left(\frac{\gamma^{2}H^{3}}{H'^{2}}\right)_{\varphi_{\mathrm{in}}}
-\left(\frac{\gamma^{2}H^{3}}{H'^{2}}\right)_{\varphi_\cl}\right]\,,\label{eq:df2}
\end{eqnarray}
respectively. [The results of Ref.~\cite{Martin:2005ir} are reproduced for $\gamma\equiv1$ in Eqs.~(\ref{eq:df1},\ref{eq:df2}).] Evidently, the only input needed to carry out the integral (\ref{eq:df1}) are the potential and the warping, which (in the slow-roll limit) completely determine $H$ and $\gamma$. In Ref.~\cite{Lorenz:2010vf}, we considered several examples, but here we focus on the case of ``UV brane inflation'' (see below) with
$V(\varphi)=V_{0}+(m^{2}/2)\varphi^{2}$ and $T(\varphi)=\varphi^{4}/\lambda$. 
In the limit where the constant $V_{0}$ dominates, Eq.~(\ref{eq:df1}) can be integrated (taking on board that $\varphi<\mpl$ always, a consequence of the string-geometric background~\cite{Baumann:2006cd}), giving\begin{equation}\label{eq:UVcase}
\mean{\df_{1}^{2}}_{_{\rm UV}} \simeq  \frac{16}{15 \sqrt{2}}
\left(\frac{m}{\mpl}\right)^2
\frac{\beta ^{3/2}}{\alpha ^{1/2}}\frac{\mpl^3}{\varphi _\cl}\, ,\quad\mean{\df_{2}}_{_{\rm UV}} \simeq  -\frac{4}{15 \sqrt{2}}
\left(\frac{m}{\mpl}\right)^2
\frac{\beta ^{3/2}}{\alpha ^{1/2}}\frac{\mpl^3}{\varphi_\cl ^2}\, ,
\end{equation}
where $\alpha \equiv 12\pi \mpl ^2/(\lambda m^2)$ and $\beta \equiv V_0/(m^2\mpl^2)$ are parameter combinations. The definition of the inflaton's Gaussian PDF (\ref{eq:Pc}) remains unchanged in the DBI case. Hence, for a given $\alpha,\beta,m$ and $\varphi_\ini$, Eqs.~(\ref{eq:UVcase}) give the mean $\mean{\varphi}=\varphi_\cl+\mean{\delta\varphi_{2}}$ for any $\varphi_\cl$ on the DBI slow-roll trajectory.

\section{Geometric consistency in compact extra dimensions}
In the 10d string theory picture, the scenario of UV brane inflation corresponds to a D3-brane inside a 6d ``warped throat'' (\ie cone-shaped) geometry, where a radial direction $r$ extends above a 5d angular base. Up to renormalization, the D3's position along $r$ is the inflaton, which therefore has a limited range $\phi\in[\phi_{0},\phi_{_{\mathrm{UV}}}]$, with $\phi_{0}$ the ``tip'' of the cone and $\phi_{_{\mathrm{UV}}}$ at its base (where it is joined onto a more complicated compact 6d geometry). Classically, the inflaton value $\varphi_\cl$ in this scenario decreases during inflation (\ie the D3 starts out close to the UV end of the throat, moving towards the tip). What, however, follows for the quantum-corrected $\mean{\varphi}$? According to Eq.~(\ref{eq:UVcase}), $\mean{\delta\varphi_{2}}$ is negative and since $\mean{\varphi}=\varphi_\cl+\mean{\delta\varphi_{2}}$, this bears the imminent risk of $\mean{\varphi}$ decreasing below $\phi_{0}$. Translated back into the D3's  $r$-coordinate, the quantum jumps hence seem to take the brane ``outside of spacetime''. While a similar effect was observed in Ref.~\cite{Martin:2005ir}, where the inflaton developed a non-zero probability for negative field values under the influence of the noise, the geometric origin of a stringy inflaton renders the issue far more severe.

\begin{figure}[t]
\begin{minipage}{0.4\textwidth}
\includegraphics[width=\textwidth]{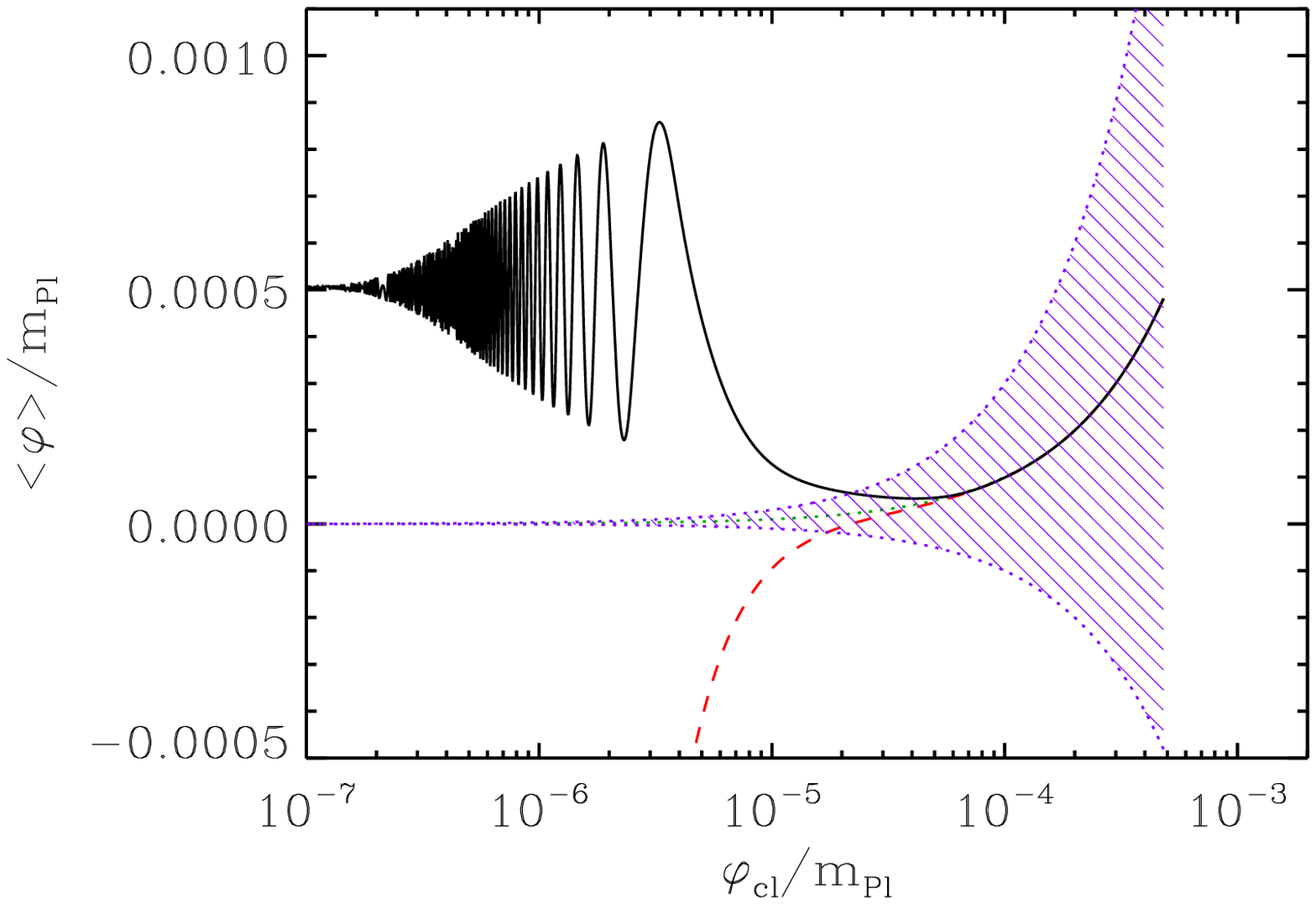}%
\end{minipage}
\begin{minipage}{0.6\textwidth}\caption{\label{fig}Evolution of a DBI inflaton for the shape of $V$ and $T$ cited in the text and $\alpha =38$, $\beta  =3.7$, $m\simeq 2.19\times 10^{-7}\mpl$ and $\varphi_\ini=5\times
  10^{-4}\mpl$.
The green dotted
  line is the \emph{classical} trajectory, 
  while the red dashed line is the \emph{quantum} mean $\mean{\varphi}=\varphi_\cl +\mean{\df_{2}}$. The black solid line corresponds to 
  $\mean{\varphi}_{\mathrm{w}}$ for absorbing walls at $\phi_0=10^{-5}\mpl$ and $\phi_{_{\rm
      UV}}=10^{-3}\mpl$. The hatched area is the validity regime of the
  perturbative treatment.

  }
\end{minipage}
\end{figure}

As advocated in Ref.~\cite{Lorenz:2010vf}, this problem is avoided if reflecting or absorbing walls are installed at certain positions in $\varphi$-space. While Ref.~\cite{Chandrasekhar:1943ws} considered the case of one reflecting wall, we derived the PDF for a stochastic (DBI or standard) inflaton field between two absorbing walls at $\phi_{0}$ and $\phi_{_{\mathrm{UV}}}$. (For UV brane inflation, this is interpreted as the D3 being ``trapped'' inside the warped throat.) To this end, we repeatedly applied the familiar method of images from electrodynamics, introducing an infinite number of ``image branes'' of alternating charge. The final PDF for this configuration is $P_{\mathrm{w}}(\varphi)=\left(2\pi \mean{\df_{1}^2}\right)^{-1/2}\tilde{P}(\varphi)$, where
\begin{equation*}
\tilde{P}(\varphi)\!=\!\!\!\sum _{n=-\infty}^{+\infty}\!\!\!\left(
\exp\left\{-\frac{\left[\varphi-\phi_{\rm m}
+2n\left(\phi_{_{\rm UV}}-\phi_0\right)\right]^{2}}
{2\mean{\df_1^2}}\right\}
-\exp\left\{-\frac{\left[\varphi+\phi_{\rm m}-2\phi_0
+2n\left(\phi_{_{\rm UV}}-\phi_0\right)\right]^{2}}
{2\mean{\df_1^2}}\right\}
\right)
\label{eq:Ptwowalls}
\end{equation*}
with $\phi_{\mathrm{m}}=\mean{\varphi}$ the former mean found with the ``no walls'' PDF~(\ref{eq:Pc}). The true mean field value in the presence of the walls then is $\mean{\varphi}_{\mathrm{w}}\propto
\int{\rm d}\varphi\,
P_{\mathrm{w}}(\varphi)\,\varphi$ and was calculated in Ref.~\cite{Lorenz:2010vf}. Interestingly, it shows oscillatory behavior, as illustrated by the example of Fig.~\ref{fig}. As the inflaton approaches the throat's tip, it seems that $\mean{\varphi}_{\mathrm{w}}$ stabilizes at a position above $\phi_{0}$. There is, however, potential concern about the validity of the perturbative approach.

\section{Conclusions}
In brane inflation, the scalar field $\phi$ corresponds to the position of a $D$-brane within the 6d compact string geometry. On the one hand, this enriches the inflationary phenomenology due to the brane's non-standard kinetic term. On the other hand, the inflaton's ten-dimensional origin leaves a geometric legacy to the four-dimensional effective theory: the values of $\phi$ should respect the field space limitations imposed by the finite size of the extra dimensions. Here, we generalized several results of stochastic inflation to DBI dynamics: where, in the standard case, integrals of $H(\phi)$ (and derivatives) were sufficient, the DBI case is described accurately only if the Lorentz factor $\gamma(\phi)$ is included alongside $H(\phi)$. We obtained the Gaussian probability distribution function $P_{\mathrm{c}}(\varphi)$ for the smoothed-out field $\varphi$, which allowed us to calculate the (second order noise-corrected) mean field value $\mean{\varphi}$ to be compared with the classical one $\varphi_\cl$.

We then argued that, to avoid conflict with the underlying 10d string theory picture, even the quantum jumps superimposed on the inflaton's DBI trajectory should not take $\varphi$ out of the geometrically admitted range. We proposed to implement this restriction by using suitable PDF boundary conditions in field space, \ie by installing walls where the compact dimension associated with $\varphi$ ``ends''. As an example, we considered a quadratic inflaton potential  together with a quartic warping and imposed absorbing boundary conditions at both ends of the (radial direction of the) extra dimensions. As seen in Fig.~\ref{fig}, the trajectory then changes significantly for a chosen parameter set. Our hope is that, as the shape of the warping and the potential are better understood from string theory, these results may shed light on the possibly complicated path of the brane (that is, the inflaton) inside the six-dimensional compact space.


\section*{References}
\bibliography{DBIbibliography}

\providecommand{\newblock}{}
\begin{thebibliography}{10}
\expandafter\ifx\csname url\endcsname\relax
  \def\url#1{{\tt #1}}\fi
\expandafter\ifx\csname urlprefix\endcsname\relax\def\urlprefix{URL }\fi
\providecommand{\eprint}[2][]{\url{#2}}

\bibitem{Dvali:1998pa}
Dvali G~R and Tye S~H~H 1999 {\em Phys. Lett.\/} {\bf B450} 72--82
  (\textit{Preprint} \eprint{hep-ph/9812483})

\bibitem{Kachru:2003sx}
Kachru S {\em et~al.\/} 2003 {\em JCAP\/} {\bf 0310} 013 (\textit{Preprint}
  \eprint{hep-th/0308055})

\bibitem{Garriga:1999vw}
Garriga J and Mukhanov V~F 1999 {\em Phys. Lett.\/} {\bf B458} 219--225
  (\textit{Preprint} \eprint{hep-th/9904176})

\bibitem{Starobinsky:1994bd}
Starobinsky A~A and Yokoyama J 1994 {\em Phys. Rev.\/} {\bf D50} 6357--6368
  (\textit{Preprint} \eprint{astro-ph/9407016})

\bibitem{Lorenz:2010vf}
Lorenz L, Martin J and Yokoyama J 2010 {\em Phys. Rev.\/} {\bf D82} 023515
  (\textit{Preprint} \eprint{1004.3734})

\bibitem{Martin:2005ir}
Martin J and Musso M 2006 {\em Phys. Rev.\/} {\bf D73} 043516
  (\textit{Preprint} \eprint{hep-th/0511214})

\bibitem{Martin:2005hb}
Martin J and Musso M 2006 {\em Phys. Rev.\/} {\bf D73} 043517
  (\textit{Preprint} \eprint{hep-th/0511292})

\bibitem{Alishahiha:2004eh}
Alishahiha M, Silverstein E and Tong D 2004 {\em Phys. Rev.\/} {\bf D70} 123505
  (\textit{Preprint} \eprint{hep-th/0404084})

\bibitem{Baumann:2006cd}
Baumann D and McAllister L 2007 {\em Phys. Rev.\/} {\bf D75} 123508
  (\textit{Preprint} \eprint{hep-th/0610285})

\bibitem{Chandrasekhar:1943ws}
Chandrasekhar S 1943 {\em Rev. Mod. Phys.\/} {\bf 15} 1--89

\end{thebibliography}

\end{document}